\documentclass{iopart}
\usepackage{iopams}  

\usepackage{graphicx}

\begin{document}

\title[Disorder effect on spin susceptibility of 2D1V electron gas]{Disorder effect on the spin susceptibility of the two--dimensional one--valley electron gas}

\author{S De Palo$^{1,2}$, S Moroni$^{1,3}$ and G Senatore$^{1,2}$}

\address{$^1$ INFM DEMOCRITOS National Simulation Center, Trieste, Italy}
\address{$^2$ Dipartimento di Fisica Teorica, Universit\`a di Trieste, Strada Costiera 11, 34014 Trieste,Italy}
\address{$^3$ SISSA, International School for Advanced Studies, via Beirut 2-4, 34014 Trieste, Italy}
\ead{depalo@democritos.sissa.it}

\begin{abstract}
  Starting from the Quantum Monte Carlo (QMC) prediction for the
  ground state energy of a clean two--dimensional one--valley (2D1V)
  electron gas, we estimate the energy correction due to scattering
  sources present in actual devices such as AlAs quantum wells and
  GaAs heterostructures.  We find that the effect of uncorrelated
  disorder, in the lowest (second) order in perturbation theory, is to
  enhance the spin susceptibility leading to its eventual divergence.
  In the density region where the Born approximation is able to
  reproduce the experimental mobility, the prediction for the spin
  susceptibility yielded by perturbation theory is in very good
  agreement with the available experimental evidence.
\end{abstract}

%\maketitle

\section{Introduction}

The two--dimensional electron gas that can be realized in quantum
wells or at the interface of semiconducting heterostructures has
attracted a lot of interest over the years\cite{Ando,Vignale}. Such
interest has been recently renewed by the the discovery of an apparent
metallic phase which is at variance with the predictions of the
scaling theory of localization for non-interacting 2D systems at zero
magnetic field\cite{Reviews}.  The strictly two-dimensional electron
gas (2DEG) embedded in a uniform neutralizing background has been
often used to describe the physics of these
devices\cite{Ando,Vignale}. However it has been recently found that
the 2DEG model is too simple to provide a quantitative account of
experiments, which can only be achieved through the inclusion in the
model of essential device details, such as  the finite transverse
thickness\cite{thick}, the in--plane anisotropic
mass\cite{Gokmen}, the valley degeneracy present for instance in $Si$-based
devices\cite{Mariapia}, the scattering sources (disorder) which determine
the mobility\cite{thick,Mariapia}.

Here, we discuss the effect of disorder on the ground state energy and
spin susceptibility of narrow $AlAs$ Quantum Wells (QW)\cite{Vakili}
and a GaAs HIGFET\cite{Zhu}, analyzing as well the role of different
scattering sources. We stress that an accurate treatment of electron
correlation is crucial in the present approach, which is based on the
properties of the ideally--clean interacting electron gas; in
particular on its ground state energy and static response functions,
the latter being a key ingredient in the evaluation of the ground
state energy shift due to disorder.  Moreover, some of the parameters
modelling the disorder are not known from experiments and we choose to
fix them by fitting the experimental mobility within the Born
approximation. The set of parameters determined in such a way is then
used to estimate the effect of disorder on the ground state energy,
within second order perturbation theory.

In the first section we introduce the model and give some details on
our estimate of the (wavevector and spin-polarization dependent)
density--density response function. We then present our results for the mobility, obtained using the Born approximation,  in the second
section. Finally, we discuss the effect of disorder on the spin susceptibility enhancement and ground state energy  in the third section and offer some conclusions.

\section{Model and Theory}

Our starting point is the strictly 2D1V electron gas (2D1VEG), whose state  at zero temperature and  magnetic field can be fixed  by  just two
dimensionless parameters: the coupling $r_s=1/\sqrt{\pi n}a_B$ and the spin polarization $\zeta=(n_{\uparrow}-n_{\downarrow})/n$.  Above, $n_{\uparrow}$ and $n_{\downarrow}$ denote the spin up and spin down areal densities, $n= n_{\uparrow}+n_{\downarrow}$, and specific parameters of the solid state device appear only in the effective Bohr radius $a_B=\hbar^2\epsilon/m_be^2$, via  the dielectric constant $\epsilon$ and the  band mass $m_b$.

In this work we assume that the ground state of the 2D1VEG in the
presence of disorder provides a first reasonable approximations to the
observed metallic phase; we assume as well that, with respect to the
ideally clean system, the ground state is not strongly altered by a
weak disorder--at least far from the metal-insulator transition--and
therefore the effect of scattering sources can be accounted for by
perturbation theory. We note in passing that a realistic description 
of these  systems must necessarily take into account disorder, in order to predict a  finite (or vanishing) mobility. 

The energy per particle of the 2D1VEG in the presence of a weak uncorrelated disorder reads, at the lowest (second) order in perturbation theory, 
\begin{eqnarray}
E(r_s,\zeta)&=&E_{2D}^{QMC}(r_s,\zeta)+\frac{1}{2n} \sum_q
\chi_{nn}(q,\zeta)\langle |U_{imp}(q)|^2\rangle_{dis} \nonumber \\ &\equiv& E_{2D}^{QMC}(r_s,\zeta)+\Delta(r_s,\zeta),
\label{diso}
\end{eqnarray}
where $U_{imp}(q)$ is the Fourier transform (FT) of the random
scattering potential and $\langle \dots \rangle_{dis}$ denotes the
average on the disorder configuration distribution. Above,
$E_{2D}^{QMC}(r_s,\zeta)$ and $\chi_{nn}(q,\zeta)$ are respectively
the energy and the density-density linear response of the ideally clean
system.  $E_{2D}^{QMC}(r_s,\zeta)$ can be readily calculated from the
analytical parametrization of Quantum Monte Carlo energies given in
Ref.~\cite{Attaccalite}. We describe how to construct  $\chi_{nn}(q,\zeta)$,  which is accurately  known only  at $\zeta=0,1$\cite{QMC_data}, in the subsection 2.1 below.    

For the extremely clean HIGFET the random scattering comes from the
unintentional doping of the $GaAs$ channel by charged impurities with density $N_d$  and/or from the charged scatterers in the $Al_{0.32} Ga_{0.68}
As$ barrier. The $U_{imp}$ for these scatterers are taken from
\cite{Gold_hete}. The unknown densities of charged scatterers ($N_d$ and
$N_{AlGaAs}$) are obtained from a fitting of the mobility as  described in Sec. 2 below. Here, we just mention  that the depletion density $N_d$ is expected to be negligible in these systems and indeed   our  best mobility fit is compatible with  $N_d=0$.

Many scattering sources contribute to the finite mobility of the QW\cite{Poortere}:
remote impurities  due to the intentional {\it
  delta} doping, three dimensional homogenous background doping with
density $N_b$ in the AlGaAs, possible unintentional doping in the AlAs channel with
density $N_c$ and above all fluctuations of the quantum well
width, which is usually modeled with a contribution to $\langle |U_{imp}(q)|^2\rangle_{dis}$   $\propto \Delta^2 \Lambda^2 e^{-q^2
  \Delta^2/4}$\cite{Gold_AlAs}. The first source can be modeled as the scattering coming from a
sheet of randomly distributed charged impurities of areal density $n_i$
separated from the side of the QW by an AlGaAs
spacer of width $d$ (in this case $n_i=5 \cdot 10^{12} cm^{-2}$ and
$d=756 A$\cite{Vakili, Poortere}).  The  unknown parameters $\Delta$, $\Lambda$,  $N_b$ and $N_c$ are fixed through the mobility fit. For completeness, we need to add that here we considered background doping only in the spacer between the QW and the delta doping sheet.

\subsection{Density-density response function}

The density-density linear response function for a partially spin polarized system
can be written in terms of local--field factors (LFF) depending on the wavevector $q$, as well as on charge and magnetization densities, respectively $n$ and $m$\cite{Vignale}:
\begin{eqnarray}
\chi_{nn}(q,\zeta) &=&\frac{\chi_0^{\uparrow}+\chi_0^{\downarrow}
+4 \chi_0^{\uparrow}\chi_0^{\downarrow} G_{mm}(q)v_{2d}(q)}{D}, \\
D=1&+&v_{2d}(q)\left[(-1+G_{mm}-2 G_{nm}+G_{nn})\chi_0^{\downarrow} \right.
\nonumber \\
   &+&   (-1+G_{mm}+2 G_{nm}+G_{nn})\chi_0^{\uparrow}
\nonumber \\
   &+& \left. 4 (-G_{mm}-G_{nm}^2+G_{mm}G_{nn})\chi_0^{\downarrow}\chi_0^{\uparrow}v_{2d}(q)\right],
\end{eqnarray}
where $\chi_0^{\sigma}(q,\zeta)$ ($\sigma=\uparrow,\downarrow$) is the spin
resolved density--density response function for the non-interacting
system\cite{Stern, Vignale}, $v_{2d}(q)$ is the Fourier transform of the Coulomb interaction, and $G_{\alpha,\beta}(q,\zeta)$ are the LFF.

\begin{figure}[tb]
\begin{center}
\includegraphics[width=13cm,angle=-0]{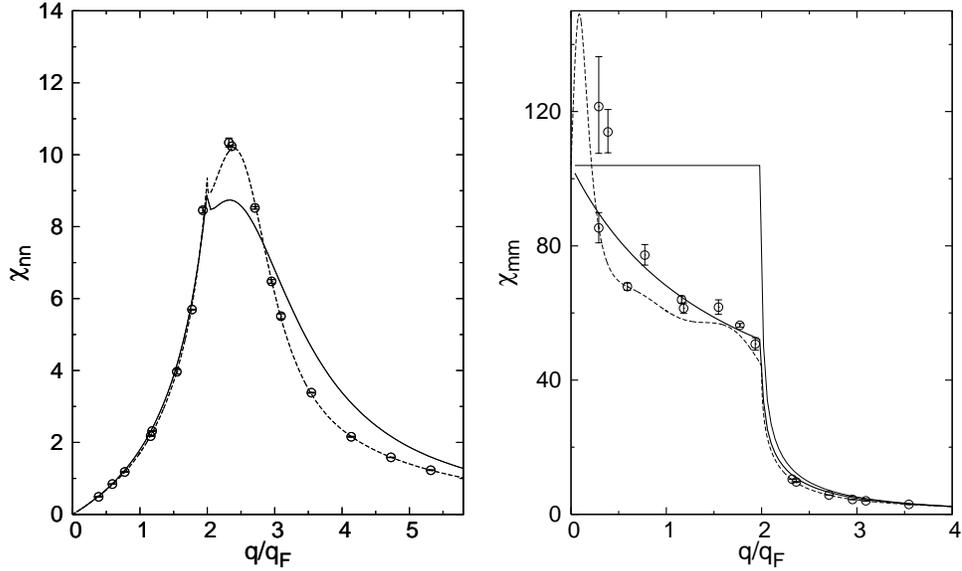}
\end{center}
\caption{Linear response of the 2DEG in $Ry^{-1}/n$. Open dots are charge ($\chi_{nn}$) and spin ($\chi_{mm}$) response functions as from QMC 
  simulations\cite{QMC_data}, the dashed lines are response functions
  using analytical parametrizations from \cite{polini}, the solid lines
  the results of the LA. For the spin case (right panel) the EA is   also shown (thick solid line). For the definition of the LA and EA see text.}
\label{risposte}
\end{figure}

A complete description of the response functions relies on the
knowledge of the LFF in the whole momentum region. We note here that the exact  low--momentum   behaviour ($q\rightarrow 0$)  of the LFF is known  in   terms of  the exchange-correlation energy
$\epsilon_{xc}$\cite{Vignale}: 
\begin{eqnarray}
\label{gnn}
G_{nn}(q)&=&-\frac{1}{v_{2d}(q)}\biggl(2 \frac{\partial \epsilon_{xc}}
{\partial n}+n \frac{\partial^2 n \epsilon_{xc}}{\partial n^2} \biggr), \\
\label{gnm}
G_{nm}(q)&=&-\frac{1}{v_{2d}(q)}\biggl( \frac{\partial \epsilon_{xc}}{\partial m}+n \frac{\partial^2 \epsilon_{xc}}{\partial n \partial m}\biggr), \\
\label{gmm}
G_{mm}(q)&=&-\frac{1}{v_{2d}(q)}\biggl(n \frac{\partial^2 \epsilon_{xc}}{\partial m^2} \biggr),
\end{eqnarray}
with $m=n_\uparrow-n_\downarrow=n\zeta$.  The simplest approximation
would be to extend the low-momenta linear behaviour
($v_{2d}^{-1}\propto q$) of the LFF to all momenta. We have tested the
effect on the response of the linear approximation (LA) for $G_{nn}$
and $G_{mm}$, at zero polarisation, with available QMC
data\cite{QMC_data}. Deviations from the LA become more evident as the
system becomes more strongly interacting.  We report in
\Fref{risposte} results for $\chi_{nn}$ and $\chi_{mm}$ at $r_s=10$
and $\zeta=0$ where the deviation of the LA from the QMC data is more
evident. Note that for $\zeta=0 $, $\chi_{nn}$ ($\chi_{mm}$ ) only
involves $G_{nn}$ ( $G_{mm}$).  For the charge case (left panel in
\Fref{risposte}) the LA for $G_{nn}$ works quite well at least up to
$2 k_F$.  For the spin case (right panel in \Fref{risposte}) the
$\chi_{mm}$ obtained from the LA for $G_{mm}$ shows important
deviations from the QMC results over the whole range $0\le q\le 2q_F$,
while a much better and in fact satisfactory agreement is obtained
with the exponential approximation, whereby
$G^{EA}_{mm}(q,\zeta)=G^{LA}_{mm}(q,\zeta)\times \exp{[-\alpha
  (q/q_F)]}$ and $\alpha=0.1$. An analytical parametrizations of the LFF\cite{polini}  embodying the exact known behaviour at small and large momenta  is available for $\zeta=0$ and $0\le r_s \le 10$. However, while it could be used for the calculation of mobility (see below)  it is of no use for the calculation of the spin susceptibility, which  requires the $\zeta$--dependence of the LFF.

In computing the $\zeta$--dependent  correction to the ground state energy, we have used the LA   for   $G_{nn}$ and $G_{nm}$ and the EA for $G_{mm}$. We do not discuss in detail here the behavior of $G_{nm}$, which appears to be one order of magnitude smaller that the other two LFF and therefore should not affect results in an appreciable manner.  

\section{The mobility}
\label{fitting}
Quite generally, not all parameters entering the modelling of the scattering sources are know from experiments and we take the customary approach in which the unkwon once are fixed through  a global fit of the  experimental mobility.   The relaxation time $\tau$ at the lowest order in the
scattering potential is given by the Born
approximation\cite{Stern_Howard_67}:
\begin{equation} \frac{1}{\tau}=\frac{\hbar^{-1}}{2\pi \epsilon_F}
\int_{0}^{2 k_F} dq\frac{q^2}{(4k_F^2-q^2)^{1/2}} \frac{\langle
|U_{imp}(q)|^2 \rangle_{dis}}{\epsilon_P(q)^2}
\label{mobi}
\end{equation} where
$\epsilon_P(q)=1-v_{c}(q)(1-G_{nn}(q))\chi_0(q)=\epsilon_P^{RPA}(q)+v_c(q)G_{nn}(q)\chi_0(q)$.
The integrand in \Eref{mobi} is peaked around $2 k_F$ because of the combined effect of the  factors $(4k_F^2-q^2)^{-1/2}$ and $\epsilon_P(q)^{-2}$, the latter being strongly enhanced by $G_{nn}(q)$ with respect to its RPA expression. An accurate  estimate of $G_{nn}(q)$    in this region of momenta  is
therefore crucial: the disorder parameters  can
increase by almost an order of magnitude if one  replace
$\epsilon_P(q)$ with $\epsilon_P^{RPA}(q)$ in the mobility fit.

\begin{figure}[tb]
\begin{center}
\includegraphics[width=14cm,angle=-0]{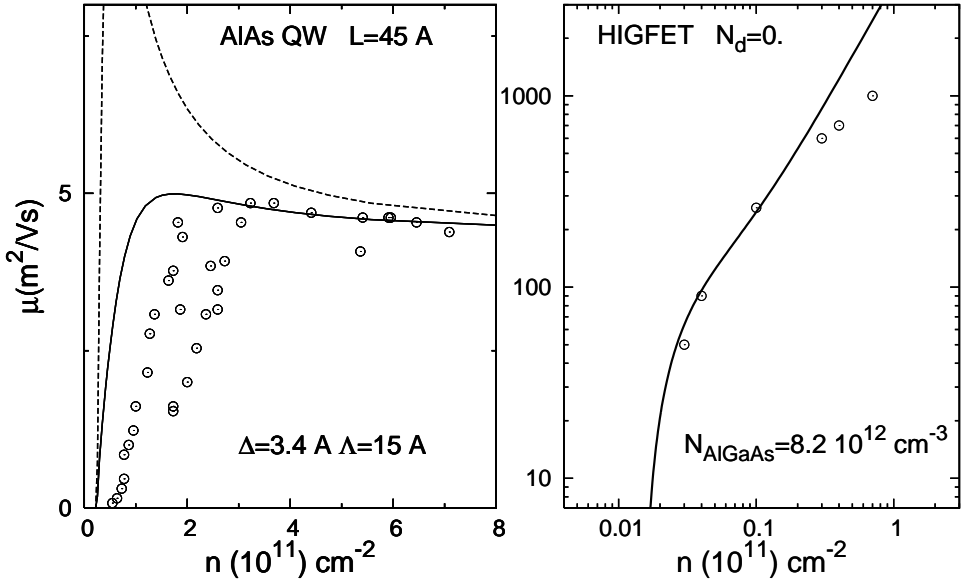}
\end{center}
\caption{ 2DEG mobility in actual devices. Open dots are experimental data for the AlAs QWs of
\cite{Vakili} (left panel) and for the HIGFET
heterostructure\cite{Zhu} (right panel). Solid lines are the fitted mobility using
\Eref{mobi}. For the QWs the mobility obtained including  only  roughness scattering is also shown (dashed line).} \label{mobility}
\end{figure}

Here  we  use  $G_{nn}$  for a strictly two dimensional system 
and accordingly we set  $v_c(q)=v_{c,2D}(q)=2\pi e^2/\epsilon q$.  This may look at first  a very crude  assumption for the HIGFET\cite{Zhu}, which is characterized by a sizeable thickness. However we have checked, within RPA, that while the fitted  disorder parameters change appreciably in going from the  $v_{c,2D}(q)$ to  $v_{c,thick}(q)$, the energy shift due to  disorder does not change sensibly provided the same consistent combination of  $v_{c}(q)$ and disorder parameters    used  for the mobility is also used for the  energy shift calculation. The same applies to the ensuing spin susceptibility.

Mobility results for the two devices  considered are shown in
\Fref{mobility}.  In the QW the surface roughness plays the major role in
determing the mobility at high densities ($\Delta=3.4 A$, $\Lambda=15
A$) in agreement with existing literature\cite{Gold_AlAs} (See left
panel of \Fref{mobility}).  At low density, however,  roughly below $n \simeq
2.5 \cdot 10^{11} cm^{-2}$, the Born approximation is not able to reproduce the  experimental data anymore.  This is a density  region where the charged
impurities ($N_b=N_c=2 \cdot 10^{14} cm^{-3}$)  become effective.  In
the right panel of \Fref{mobility} we display results for the 
HIGHFET with $N_d=0$ and $N_{AlGaAS}=8.2 10^{12} cm^{-3}$. The discrepancy of  these disorder parameters with those in \cite{thick} is due to the replacement, with respect to the previous  calculation,  of $v_{c,thick}$ with $v_{c,2D}$. The effect of such a change on the spin susceptibility is however barely visible, as it can be checked by comparing the results in \Fref{chis}  with those in \cite{thick}. We should mention that both in the present calculations and those of \cite{thick} we have chosen the form of $U_{imp}$ appropriate to  a thick electron gas.   

\section{The spin susceptibility} The spin susceptibility enhancement of the systems under investigation is\cite{thick}
\begin{equation} \frac{\chi_s}{\chi_0}=\biggl[ \frac{\partial^2 E_0
(r_s,\zeta)}{\partial \zeta^2}\biggr]_{\zeta=0} \biggl[
\frac{\partial^2 E^{QMC} (r_s,\zeta)}{\partial \zeta^2}
+\frac{\partial^2\Delta E (r_s,\zeta)}{\partial \zeta^2}
\biggr]_{\zeta=0}^{-1},
\label{chi_spin}
\end{equation} where $E_0(r_s, \zeta)$ is the energy of the
non-interacting system,   $\Delta E
(r_s,\zeta)$  the energy shift due to disorder defined in \Eref{diso}, and $E^{QMC} (r_s,\zeta)$  the energy of the clean system, which may include,  
if necessary, the effect of thickness. As mentioned above, the results of this calculation strongly depend on the LFF  in the region around $2q_F$, region where  $\chi_{nn}(q,\zeta)$ has a sizeable change when varying $\zeta$ around zero.  We stress that the parameters describing the disorder are fixed by a fit of the experimental mobility and depend on  $G_{nn}$ at zero
polarisation; while $\Delta E (r_s,\zeta)$ requires the knowledge of
all LFF and their $\zeta$--dependence.  

\begin{figure}[tb]
\begin{center}
\includegraphics[width=14cm,angle=-0,trim=0 8 0 0]{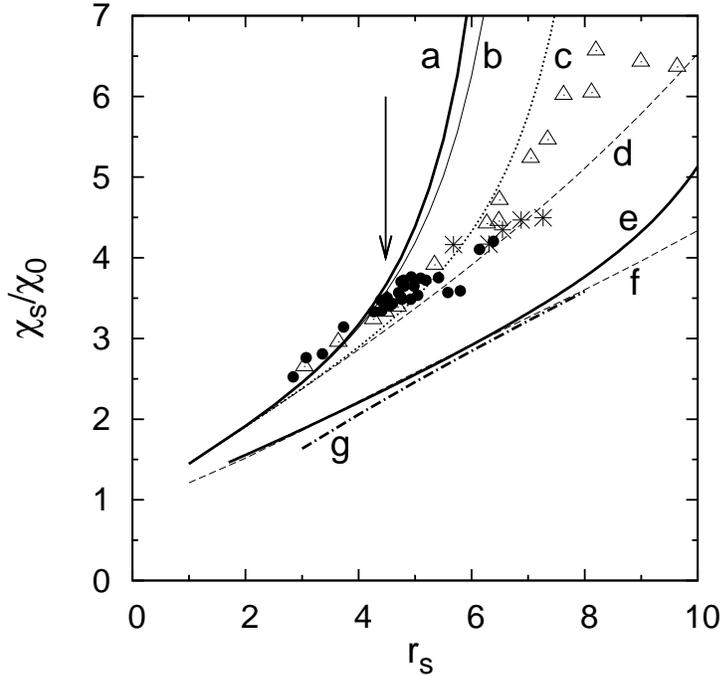}
\end{center}
\caption{Spin susceptibility enhancement of the 2DEG in actual devices. Experimental results  for the thin
QW of  \cite{Vakili} and for  the HIGHFET of  \cite{Zhu}  are respectively represented by
points (different  symbols correspond to different samples\cite{Vakili}) and by the line labeled {\bf g}.  
Line {\bf d} reports the QMC  prediction for the clean 2DEG\cite{Attaccalite}, while   lines  {\bf f} and  {\bf e}   give the QMC based
predictions for the {\it clean}  and {\it dirty} quasi-2DEG in the HIGHFET\cite{thick}, respectively.  The arrow indicates the density at which the Born
approximation for the QW mobility fails.   Line {\bf a} is our
prediction for $\chi_s/\chi_0$ in the QW  including both  surface roughness  and charged impurities scattering; the predictions obtained including only surface roughness or only charged impurities are given by lines 
 {\bf b} and {\bf c}, respectively.}
\label{chis}
\end{figure}

Before examining in detail our results for the spin susceptibility,
summarized in \Fref{chis}, we make some general comments on the effect
of disorder. Apart from the surface roughness at very high density
($r_s\le 1$), where it induces a negligible reduction of spin
susceptibility, the effect of all scattering sources is to enhance
$\chi_s/\chi_0$, once electron correlation is included in the response
function, even at RPA level.  A calculation employing the response
function  $\chi_0(q,\zeta)$ of  non-interacting electrons and
including only the roughness scattering, for example, predicts a
suppression of the spin susceptibility at all densities.  On lowering
the electron density the relative contribution of disorder to the second
derivative of the energy, with respect to $\zeta$, increases in size
and being negative leads to the eventual divergence of the spin
susceptibility.  We note that quite generally the transverse thickness
reduces the spin susceptibility of a 2D electron systems, while
disorder generally enhances it\cite{thick}.

As it is clearly seen in \Fref{chis}, in the extremely clean case of the
HIGHFET, the inclusion of disorder does not alter the agreement
between the theoretical prediction (obtained including thickness) and
measurements, throughout the whole experimental density
range\cite{nota}. If one neglects thickness and uses the disorder
parameters fitted to the experimental mobility of the HIGHFET, as
specified above, the energy shift due to disorder makes the
ferromagnetic state of the strictly 2DEG energetically favourable with
respect to the normal state at $r_s\simeq 12.5$. 

In contrast, the same procedure using the disorder parameters appropriate to the thin electron gas
realized in  AlAs QWs\cite{Vakili}  predicts  a transition towards a partially polarised state at 
$r_s\simeq 7$, namely a second-order phase transition.  We should stress, however, that for this system 
the fitting of the experimental  mobility in the Born approximation breaks down at low densities 
(corresponding to  $r_s \gtrsim 4$), as clearly  shown in \Fref{mobility}. 
Yet, up to $r_s \lesssim 4$, our prediction for the  spin susceptibility  is only moderately affected by disorder (thick solid line (a) in \Fref{chis}), with an enhancement with respect to the clean system of at most 20\%, which  results in a very good agreement   with  experiments.
By looking at the theoretical prediction for the spin susceptibility
enhancement obtained including only  the scattering  by charged
impurities (dotted line (c)) or only that by roughness (thin solid line (b)), 
it is evident the major role played by roughness at all densities, as well as the negligible 
effect of charged scatterers at high density (due to screening).  
At low densities, though being quite smaller than that of roughness,  
the effect of  charged impurities on $\chi_s$ becomes however sizeable.
Within second order perturbation theory, lowering the density, disorder becomes more and more effective
enhancing the spin susceptibility and finally driving it to diverge. A strong enhancement is found 
also in the experiments, however we cannot push our quantitative comparison between the experiments and 
our predictions in density regions where the level of disorder 
cannot be reliably related to the experimental mobility using the Born approximation.

We stress that the accuracy of the prediction of the spin
susceptibility of the clean 2D1V electron gas is crucial in the
present approach, as suggested by the comparison between theory and
experiment for the thin electron gas realized in narrow $AlAs$
QWs\cite{Vakili}, for which the effect of thickness is
negligible\cite{thick}. In this respect, we recall that RPA predicts
for the 2D1V electron gas a first order ferromagnetic transition
already at $r_s \simeq 5.5$ and a $\chi_s/\chi_0$ divergence in the paramagnetic phase at $r_s
\simeq 7.3$ \cite{DasSarma}. Evidently the inclusion of disorder in RPA
would push the Bloch and  Stoner transitions\cite{DasSarma} at higher density, well
inside the experimental range.

\section{Conclusions}

We have studied the effect of disorder on the spin susceptibility of 2D electron systems realized in semiconductor heterostructures, considering 
narrow $AlAs$-based QWs and  a $GaAs$-based HIGHFET, systems which have an in-plane isotropic mass and no valley degeneracy. We take as reference, in assessing the effect of disorder, the ideally clean 2D1V electron gas, whose spin susceptibility is known with  great accuracy, thanks to QMC simulations\cite{Attaccalite}.  We found that the effect of a weak  uncorrelated disorder is to enhance the spin susceptibility,  at the lowest order in perturbation theory, with correlation seemingly  playing a crucial
role.  The disorder parameters which were not known from  experiments were determined through a fit of the experimental mobility over the whole experimental density range, in the Born approximation, and then used without any change in the spin susceptibility calculation. We discovered that, at densities where  the Born approximation is capable of fitting the experimental mobility, also our prediction of  the spin susceptibility in the {\it dirty} system turns out to be  very accurate; while it appreciably deviates  from the experiment at densities where the Born approximation breaks down, or more precisely, is unable to fit the experimental mobility. 

Thus, the really weak disorder present in the $GaAs$ HIGFET of
\cite{Zhu} has a small effect on the spin susceptibility and does not change
qualitatively the phase diagram of the 2D1V electron gas. Evidently, the disorder in the $AlAs$ QWs 
of \cite{Vakili} is much stronger and it can be possibly treated  in perturbation theory only at 
densities not too low. It is anyhow reassuring that when the perturbative approach is capable of quantitatively fitting the experimental mobility also the resulting prediction of the spin susceptibility enhancement is in good agreement with experiments.


\begin{thebibliography}{90}
\bibitem{nota} The apparent discrepancy between theory end experiment at high density  (one would expect $\chi_s/\chi_0$ to go to 1 as $r_s\rightarrow 0$, while the  expression fitted  to experiments  in \cite{Zhu} tends to 0 in this limit) is eliminated once band structure effects, modifying the band mass and g-factor, are duly taken into account\cite{tan}.   We do not report here the data of \cite{tan} for $\chi_s/\chi_0$, which are obtained from a different sample for which we do not know  all relevant physical parameters.
\end{thebibliography}
\end{document}